\documentclass[12pt]{article}
\usepackage{amsmath}
\usepackage{graphicx}
\usepackage{float}
\usepackage{natbib}
\usepackage{url} 

\newcommand{\blind}{0}

\usepackage{caption}
\usepackage{subcaption}
\begin{document}

\def\spacingset#1{\renewcommand{\baselinestretch}%
{#1}\small\normalsize} \spacingset{1}


\if0\blind
{
  \title{\bf Visualizing Variable Importance and Variable Interaction Effects in Machine Learning Models}
  \author{Alan Inglis\thanks{
    Alan Inglis and Andrew Parnell's work was supported by a Science Foundation Ireland Career Development Award grant 17/CDA/4695. In addition Andrew Parnell’s work was supported by: an investigator award (16/IA/4520); a Marine Research Programme funded by the Irish Government, co-financed by the European Regional Development Fund (Grant-Aid Agreement No. PBA/CC/18/01); European Union’s Horizon 2020 research and innovation programme InnoVar under grant agreement No 818144; SFI Centre for Research Training in Foundations of Data Science 18CRT/6049, and SFI Research Centre awards I-Form 16/RC/3872 and Insight 12/RC/2289\_P2. For the purpose of Open Access, the author has applied a CC BY public copyright licence to any Author Accepted Manuscript version arising from this submission.}\hspace{.2cm} \\
    Hamilton Institute, Maynooth University\\
    and \\
     Andrew Parnell$^*$\\
    Hamilton Institute, Insight Centre for Data Analytics, Maynooth University\\
    and\\
    Catherine B. Hurley \\
    Department of Mathematics and Statistics, Maynooth University
    }
  \maketitle
} \fi

\if1\blind
{
  \bigskip
  \bigskip
  \bigskip
  \begin{center}
    {\LARGE\bf Title}
\end{center}
  \medskip
} \fi

\bigskip

\begin{abstract}
Variable importance, interaction measures, and partial dependence plots are important summaries in the interpretation of statistical and machine learning models. In this paper we describe new visualization techniques for exploring these model summaries. We construct heatmap and graph-based displays showing variable importance and interaction jointly, which are carefully designed to highlight important aspects of the fit. We describe a new matrix-type layout showing all single and bivariate partial dependence plots, and an alternative layout based on graph Eulerians focusing on key subsets. Our new visualizations are model-agnostic and are applicable to regression and classification supervised learning settings. They enhance interpretation even in situations where the number of variables is large. Our R package \texttt{vivid} (variable importance and variable interaction displays) provides an implementation.
\end{abstract}

\noindent%
\vfill

\newpage
\spacingset{1.5} 

\section{Introduction}
\label{sec:intro}
Visualization is a key tool in understanding statistical and machine learning models. 
In this paper we present new visualizations to serve two main goals, namely improved model understanding and interpretation. Our new visualizations are based on variable\footnote{We use the term `variable' throughout to denote the input to a statistical and machine learning model as this seems to be the most common parlance. Other terms commonly used include: feature, predictor, explanatory variable, independent variable, etc} importance and interaction measures, and partial dependence plots. 
A variable importance value is used to express (in a scalar quantity) the degree to which a variable affects the response value through the chosen model. A variable interaction is a scalar quantity that measures the degree to which two (or more) variables combine to affect the response variable. Variable importance and variable interaction (henceforth VImp and VInt; together VIVI) are widely used in many fields to understand and explain the behaviour of a model. In biology they are used to examine gene-gene interactions \citep[e.g.][]{IntBio}. In high-energy physics VImp can be an important tool in high dimensional feature selection processes \citep[e.g.][]{physics}. In econometrics they are common tools to evaluate interaction behaviour \citep[e.g.][]{econ}.

Traditional methods of displaying VImp or VInt use variants of line or bar plots, see for example \citet{molnar2019}. However, in variable importance plots there is relatively little emphasis on displaying how
pairs of interacting variables may be important in a model.
 This can be a hindrance to model interpretation, especially if a variable has low importance but a high interaction strength. The inclusion of interacting terms in a model has been shown to affect the prediction performance \citep{Int_Perform}. However, as shown in \cite{HighIntLowImp}, for high-dimensional models that are governed mainly by interaction effects, the performance of certain types of permutation-based variable importance measures will decrease and thereby produce low values of importance. Consequently, viewing the VInt and VImp together provides a more complete picture of the behaviour of a model fit. 
  
Our new displays present VInt and VImp jointly in a single plot. We allow for seriation so that variables are reordered with those exhibiting high VIVI grouped together. This assists in interpretation and is particularly useful as the number of variables becomes large. Furthermore, we make use of filtering, so less influential variables can be removed. For our network displays we use graph clustering to group together  interacting variables.

Partial dependence plots (PDP) were introduced by \citet{Friedpdp} to show how the model's predictions are affected by one or two predictors. In addition to the above we propose a new display which shows all pairwise partial dependence plots in a matrix-type layout, with a univariate partial dependence plot on the diagonal, similar to a scatterplot matrix. With this display the analyst can explore, at a glance, how important pairs of variables impact the fit. Once again, careful reordering of the variables facilitates interpretation.

Our final display takes the filtering of all pairwise partial dependence plots a step further. We select only those pairwise partial dependence plots with high VInt, and display an Eulerian path visiting these plots by extending the zigzag display algorithm of \cite{ZEN}. We call this a zen-partial dependence plot (ZPDP).

These new visualizations can be used to explore machine learning models more thoroughly in an easily interpretable way,  providing useful insights into variable impact on the fit. 
This is demonstrated by practical examples.
In each plot careful consideration is given to various aspects of the design, including color choices, optimising layouts via seriation, graph clustering, and Euler paths for the ZPDP. Filtering options limit the plots to variables deemed relevant from VImp or VInt scores. Our new displays are appropriate for supervised regression and classification fits, and are model and metric agnostic in that no particular model fit nor importance method is prescribed. The 
methods described here  are implemented in our R package \texttt{vivid} \citep{vivid}.

The organisation of the paper is as follows. In Section \ref{sec:vimpvint} we discuss the concepts of VImp and VInt. Then we describe our new heatmap and network displays of joint variable importance and interaction
and demonstrate these on an example.
 In Section \ref{sec:pdpVis} we discuss our new layouts for collections of partial dependence plots, either in a matrix format or zig-zag layout and show their application. 
 In Section \ref{sec:example} we use our new methodology to explore a machine learning fit from a larger dataset. Finally in Section \ref{sec:conclusion}, we offer some concluding discussion.

 \section{Visualizing variable importance and interaction}
\label{sec:vimpvint}

We begin with a non-exhaustive review of the concepts of VImp and VInt. Though the visualizations we present are agnostic to the measures used to determine these scalar quantities, some degree of understanding is helpful in interpreting the later plots.
We then describe our new visualizations and their design principles and provide illustrations.

\subsection{Measuring variable importance}
\label{sec:vimp}

A VImp is a scalar measure of a variable's influence on the response. Many techniques have been proposed to calculate variable importance, depending on the type of model. The term `influence'  here may encompass changes in the mean response or that of higher order uncertainty. In our work we focus exclusively on changes in the mean. For a wider review of variable importance techniques, and the different goals that a variety of approaches may achieve see \cite{VimpReview}.

Much of the initial work in VImp focused on estimating the partial derivative of the response with respect to one or two input variables \citep{SensAnal}. This is a global VImp measure when the model is linear, but perhaps less useful (though still potentially interesting) in non-linear models where it is often defined as a local importance measure. In high dimensional settings these methods can be discretized across a hyper-cube to allow for the identification of, e.g., linearity in a non-linear model \citep{illSA}. Due to their local behaviour, we do not incorporate them into our visualizations below. 

Some VImp measures arise naturally out of a model structure. The most familiar would be those based on summary statistics created from regression models, such as standardized coefficient values, (partial) correlation coefficients, and $R^2$. Many of these can be extended to non-linear models such as generalized additive models \citep{Smooth}, or projection pursuit regression \citep{ProjectionPR}. $R^2$ in particular seems useful as a VImp measure, as it can be defined for a wide variety of statistical models and can be decomposed into main and potentially high order interaction effects, yielding a VInt measure in addition. 

Similarly, other model-structure based methods arise out of now standard machine learning techniques. Random forests, for example, involves the use of the Gini coefficient, and the reduction in mean square error, to catalog a variable's influence on the `purity' of a model output \citep{breiman2001randomforest}. This can naturally be seen as a VImp measure. Others have extended these approaches to introduce conditional and permutation VImp statistics which aim to reduce the bias that may occur due to variable collinearity \citep[for example, see][]{RecursPart}.

Conditional variants of permutation variable importance were proposed by \cite{CondVarImp} for a random forest. This method examines splits of the trees in a random forest and permutes the variables within these subgroups (see Section \ref{sec:VarInt} for more details). Whereas \cite{CondVarImp} relied on the splitting of trees to determine the subgroups, a model-agnostic approach was introduced by \cite{CondVarImpSub} that builds the subgroups explicitly from the conditional distribution of the variables.
In tree-based models such as CART and random forests, \cite{minDepth} proposed a VImp
called minimal depth, which is the proximity of a variable to the root node, averaged across all trees.

Permutation importance was introduced by \cite{breiman2001randomforest} and is measured by calculating the change in the model's predictive performance after a variable has been permuted. The algorithm works by initially recording the model's predictive performance, then, for each variable, randomly permuting a variable and re-calculating the predictive performance on the new dataset. The variable importance score is taken to be the difference between the baseline model's performance and the permuted model's performance when a single feature value is randomly shuffled. A similar agnostic permutation concept was developed by \cite{Fisher}. This method permutes inputs to the overall model instead of permuting the inputs to each individual ensemble member. In situations where no embedded variable importance is available, a model-agnostic approach such as permutation importance is a useful tool. 

In theory any of the above global importance measures could be used in our visualizations. However, providing code for each would be a daunting task. Instead we take a pragmatic approach and use the associated VImp measure with the model that we are fitting. In cases where there is no such obvious method, we use the \cite{Fisher} agnostic permutation approach discussed above to measure VImp.\footnote{In our implementation, any available VImp may be used.}

\subsection{Measuring variable interaction}
\label{sec:VarInt}

Measuring variable interaction in a machine learning model can be considerably harder than estimating marginal importance. Even the definition of the term `interaction' is disputed \citep{letterToEditor}. We focus here on bivariate interaction only, though higher order interactions may certainly be present in many situations. \cite{Friedmans_H} state that a function $f(\boldsymbol{x})$ exhibits an interaction between two of its variables $x_k$ and $x_l$ if the difference in the value of a function $f(\boldsymbol{x})$ as a result of changing the value of $x_k$ depends on the value of $x_l$. That is, the effect of one independent variable on the response depends on the values of a second independent variable. Often, an interaction is taken to mean a simple multiplication of two (continuous) variables \citep[e.g.][]{interactionCox}, though in machine learning models much more complex relationships can exist. We follow the definition of \citet{Friedmans_H} by considering an interaction to be estimated from the difference between joint and marginal partial dependence; a full mathematical definition is given below. Even this definition should not be used without care, as in the case of highly correlated or potentially confounding variables.

In tree-based models such as CART and random forests, much focus has been on measuring interactions via the structure of trees  \citep[e.g.][]{minDepth,DTint}. If two variables are used as splits on the same branch, this might initially appear like a measure of interaction. However, this does not separate out the interaction from potential marginal effects. The problem is partially overcome by permuting the variables (individually for a VImp, jointly for VInt), to assess the effect on prediction performance. The resulting VInt measure is known as pairwise prediction permutation importance \citep{DTint2}.

For models that are not tree-based, or when a model-agnostic measure is required, a variety of other methods can be used. Many of these are based on the idea of partial dependence \citep{Friedpdp}. The partial dependence measures the change in the average predicted value as specified feature(s) vary over their marginal distribution. 
The partial dependence of  the model fit function $g$  on predictor variables $S$  (where $S$ is a subset of the $p$ predictor variables)  is estimated as:
\begin{equation}
    f_S(\boldsymbol{x}_S) = \frac{1}{n} \sum_{i=1}^{n} g(\boldsymbol{x}_S, \boldsymbol{x}_{C_i})
\end{equation}
where $C$ denotes predictors other than those in $S$, $\{\boldsymbol{x}_{C_1}, \boldsymbol{x}_{C_2},...,\boldsymbol{x}_{C_n}\}$ are the values of $\boldsymbol{x}_{C}$ occurring in the training set of $n$ observations, and $g()$ gives the predictions from the machine learning model. For one or two variables, the partial dependence functions $f_S(\boldsymbol{x}_S) $  are plotted (a so-called PDP) to display the marginal fits.

Friedman's $H$-statistic or $H$-index  \citep{Friedmans_H} is a VInt measure  created from the partial dependence by comparing the partial dependence for a pair of variables to their marginal effects. Squaring and scaling gives a value in the range $(0,1)$:
\begin{equation}
H^2_{jk} = \frac{\sum_{i=1}^{n}[f_{jk}(x_{ij}, x
_{ik}) - f_{j}(x_{ij}) - f_{k}(x_{ik})]^2 }{\sum_{i=1}^{n}f^2_{jk}(x_{ij}, x_{ik})}
\label{eqn:HStatInPaper}
\end{equation}
\noindent where $f_{j}(x_j)$ and $f_{k}(x_k)$ are the partial dependence functions of the single variables and $f_{jk}(x_j,x_k)$ is the two-way partial dependence function of both variables, where all partial dependence functions are mean-centered.

The $H$-statistic requires $O(n^2)$ predicts for each pair of variables, and so can be slow to evaluate.
Sampling from the training set will reduce the time, though at a cost of increasing the variance of the partial dependence estimates
and the $H$-statistic.

When the denominator in Equation \ref{eqn:HStatInPaper} is small, the partial dependence function for variables $j$ and $k$ is flat,
and small fluctuations in the numerator can yield spuriously high $H$-values. 
Biased partial dependence curves will also lead to inflated $H$.
This occurs in some machine learning approaches which exhibit regression to the mean in their one-way partial dependencies.
Furthermore biased partial dependence curves are a particular problem in the presence of correlated predictors. 
These issues with the $H$-statistic seem to be not widely known by practitioners \citep[though see][]{ALE}, and we provide a short illustration of these problems in the appendix. 

In our visualizations throughout this paper, we use the square-root of the average un-normalized (numerator only) version of Friedman's $H^2$ for calculating pairwise interactions:
\begin{equation}
H_{jk} = \sqrt {  \frac{1}{n} \sum_{i=1}^{n}[f_{jk}(x_{ij}, x
_{ik}) - f_{j}(x_{ij}) - f_{k}(x_{ik})]^2   }
\label{eqn:HStatInPapernn}
\end{equation}
This  reduces the identification of spurious interactions and provides results that are on the same scale as the response (for regression). It does not, however,  remove the possibility that some large $H$-values arise from correlated predictor variables.

We follow the convention of \citet{elementsStats} by using the logit scale for both the partial dependence and in calculation of the $H$-statistic when fitting a classification model with a binary response. If the response is multi-categorical a near-logit is used, defined as:
\begin{equation}
g_k(x) = \log[p_k(x)] - \frac{1}{K} \sum_{k=1}^{K}\log[p_k(x)]
\label{eq:multilogit}
\end{equation}
where $k = 1,2,...,K$ and $p_k(x)$ is the predicted probability of the $k$-th class. PDPs of $g_k(x)$ from Equation \ref{eq:multilogit} can reveal the dependence of the log-odds for the $k$-th class on different subsets of the input variables.

Alternatives to the $H$-statistic have been suggested, which could be used in place of the the $H$-statistic in our visualizations.
\citet{VIN} uses a functional ANOVA construction to   decompose the prediction function into variable interactions and main effects.
\citet{vip} suggested a partial dependence-based feature interaction which uses the variance of the partial dependence function as a measure of importance of one variable conditional on different fixed points of another.

\subsection{Heatmap visualization with seriation}
\label{sec:heatmapvis}

Traditionally, variable importance and interaction are displayed separately, with variable interaction itself spread over multiple plots, one for each variable. We direct the reader to Chapter 8 of \citet{molnar2019} for examples.
We propose a new heatmap display showing VImp on the diagonal and
VInt on the upper and lower diagonals.
The benefit of such a display is that one can see which variables are
important as individual predictors and at the same time see which pairs of variables
jointly impact on the response. It also facilitates easy comparison of multiple model fits, which is far less straightforward with separate VImp and VInt displays.

We illustrate the heatmap using a random forest fit to a college applications data set \citep{college}, with Enroll (i.e., the number of new students enrolled) as the response.
 The data was gathered from 777 colleges across the U.S. and contains 18 variables ranging from economic factors  (such as room and board and book costs) to the number of applications received and accepted. As some of the variables are skewed they are log-transformed prior to building the model. The data was split 70-30 into training and test sets. A value of $R^2 = 0.96$ was obtained for the test set. All plots were made from the training set. See the supplementary materials for a description of the data and transformations.

\begin{figure}[htbp]
\begin{center}
   \begin{subfigure}{0.49\linewidth} \centering
     \includegraphics[scale=0.4]{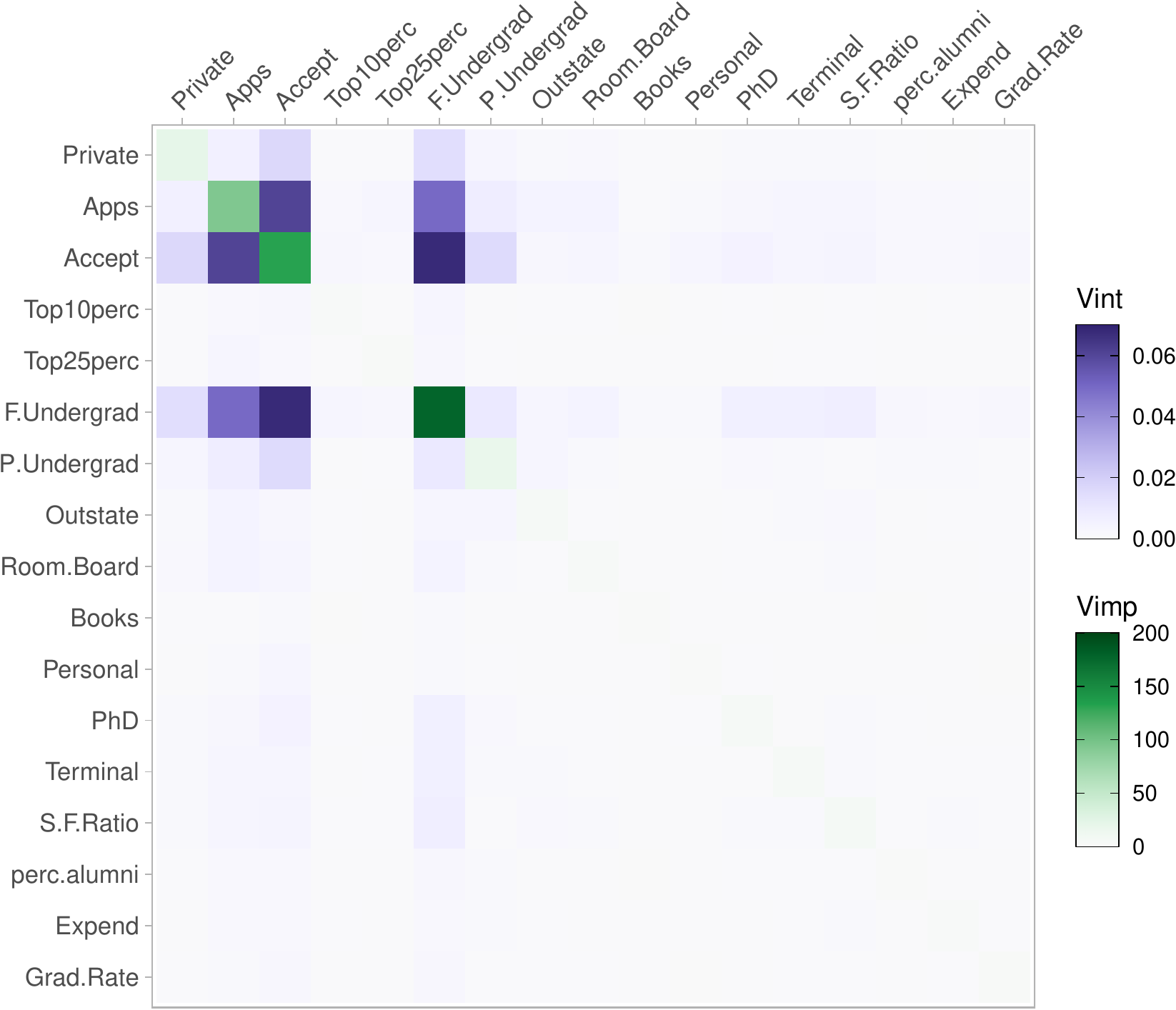} 
     \captionsetup{justification=centering}
     \caption{Unsorted ordering}
   \end{subfigure}
   \begin{subfigure}{0.49\linewidth} \centering
     \includegraphics[scale=0.4]{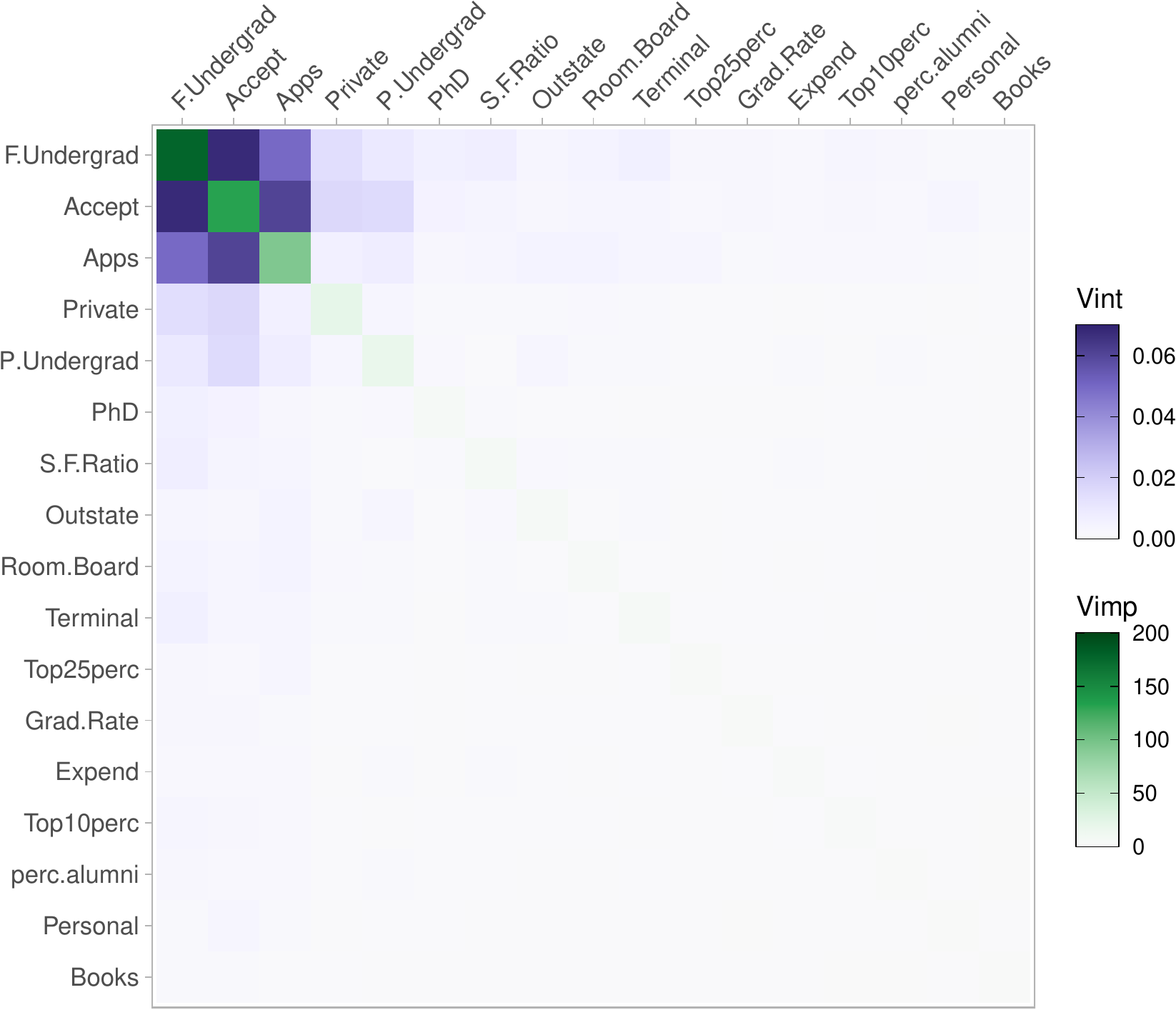} 
     \captionsetup{justification=centering}
     \caption{Leaf sort algorithm}
   \end{subfigure}
\caption{Heatmap from random forest of college application data. In (a) variables are in original order. In (b), the heatmap is re-ordered using leaf sort. In (b) we can see three important and mutually interacting variables, F.Undergrad, Accept and Apps.} 
\label{fig:dendser}
\end{center}
\end{figure}

Figure \ref{fig:dendser} 
shows our heatmap with two different orderings. Figure \ref{fig:dendser}(a) has the variables in
their original order, while Figure \ref{fig:dendser}(b) uses the leaf-sorting algorithm (described below).
The purple color scale used on the off-diagonal shows the Friedman's $H$-statistic values (un-normalized) with deeper purple indicating a higher VInt.
Similarly the green color scale on the diagonal represents the level of VImp, here measured using an embedded approach supplied by the random forest (in this case, the increase in node purity). We use colorblind-friendly, single-hued sequential color palettes from  \citet{colorspaceJSS} going from low to high luminance in both cases, designed to draw attention to high VInt/VImp variables. From the improved ordering in Figure \ref{fig:dendser}(b),  there are three clearly important and potentially interacting variables, F.Undergrad (the number of full-time undergraduate students), Accept (the number of applicants accepted), and Apps (the number of applications received), with F.Undergrad having the largest VImp when predicting Enroll. 

Many authors have investigated the benefits of re-ordering (also known as seriation) for graphical displays, see for example \citet{Hurley2004}, \citet{Seriation} and \citet{dendserPaper}. The benefits of reordering the variables in Figure \ref{fig:dendser}(b) are clear. The right-hand plot lends itself to easy interpretation whereas the left-hand plot does not.

Most seriation algorithms start with a matrix of dissimilarities or similarities between objects and produce an ordering where similar objects are nearby in the sequence.
Our goal here is a little different. As well as placing mutually interacting variables nearby in the sequence,  we would like to bring important variables or pairs of variables to the start of the sequence so that the most relevant portion of the heatmap will be in the top-left corner. 

We  use the leaf sort seriation algorithm from \citet{dendserPaper}. This uses hierarchical clustering followed by a sorting step.
Let $v_i$ be a measure of variable importance and $s_{ij}$ be the interaction measure between variables $i$ and $j$. 
Treating the matrix of interactions as a similarity matrix,
we first construct a hierarchical clustering. This  produces a  dendrogram, resulting in a variable ordering 
where high-interacting variables are nearby. Using this ordering in a heatmap generally brings high interactions close to the diagonal, but ignores our goal of placing important variables early in the sequence. 
For the sorting step we
calculate for each variable a combined measure of its importance and contribution to the interactions,  defining these scores as:
\begin{equation*}
w_i = \lambda_1 v_i + \lambda_2 \max_{j \neq i} s_{ij}.
\end{equation*}
Here $\lambda_1$ and $\lambda_2$ are scaling parameters  to account for the fact that variable importance and interaction are not measured in the same units. Reasonable choices of $\lambda_1$ and $\lambda_2$ rescale importance and interaction to, say, unit range or unit standard deviation. We
use unit range by default.
As there are many possible dendrogram orderings consistent with a hierarchical clustering of the matrix of interactions, the sorting step re-orders the dendrogram leaves so that the weights $w_i$ are generally decreasing.

Sorting the variables in this way will achieve our goals of placing high-interacting pairs of variables nearby in the sequence, while simultaneously pulling predictors with high importance and interaction to the top-left of the heatmap, leaving less relevant predictors to the bottom-right. Setting $\lambda_2 = 0$ or $\lambda_1=0$ produces plots which sort by descending VImp or max VInt respectively. For all future heatmap plots, we use the sorting strategy discussed above to optimize the arrangement of variables. After using seriation to re-order the heatmap variables, filtering can be applied to
limit the display to the most important or interacting variables; this strategy is especially useful when there are large numbers of
predictors. 

\begin{figure}[htbp]
\begin{center}
   \begin{subfigure}{0.49\linewidth} \centering
     \includegraphics[scale=0.4]{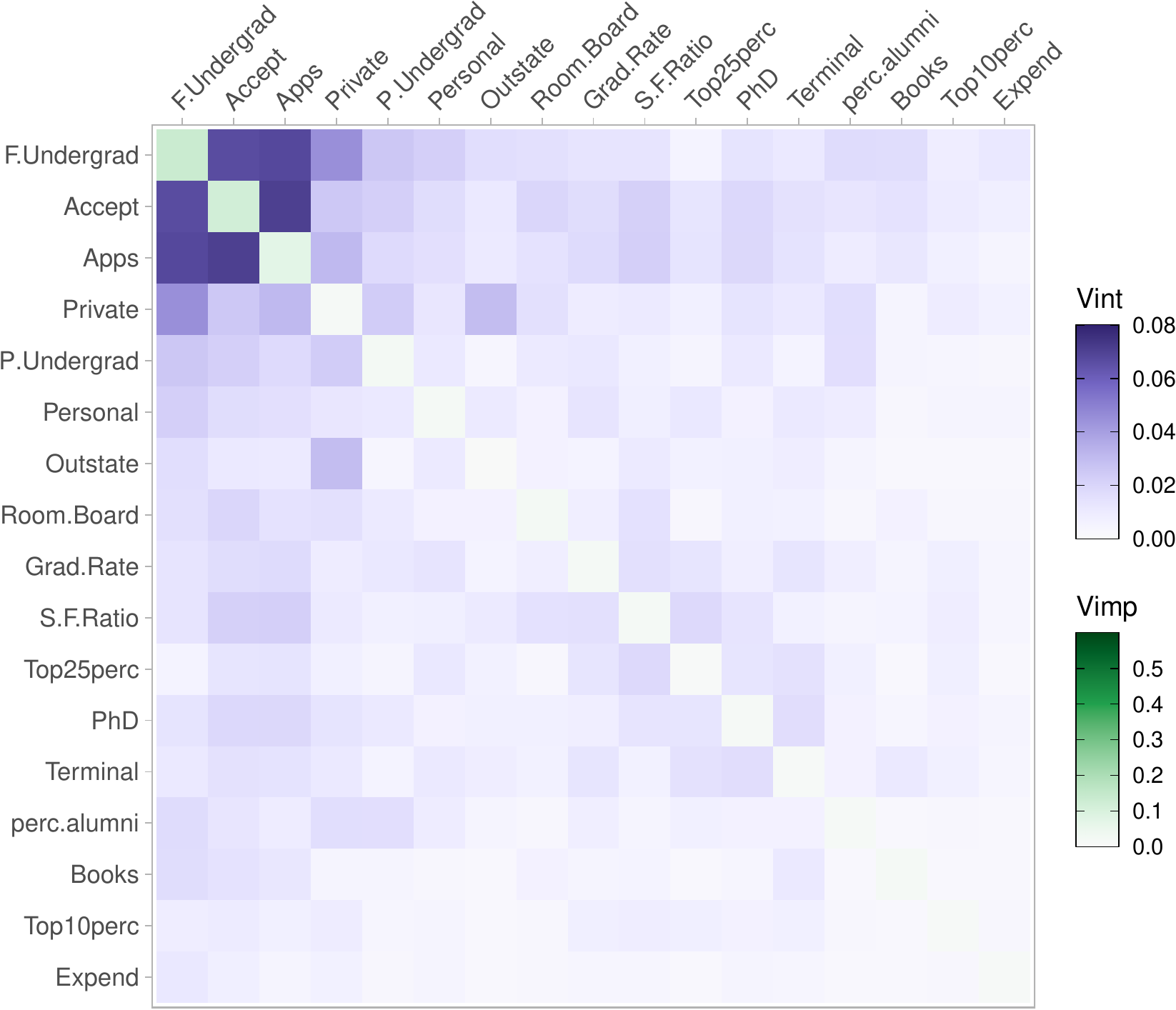} 
     \captionsetup{justification=centering}
     \caption{kNN fit}
     \label{fig:figA}
   \end{subfigure}
   \begin{subfigure}{0.49\linewidth} \centering
     \includegraphics[scale=0.4]{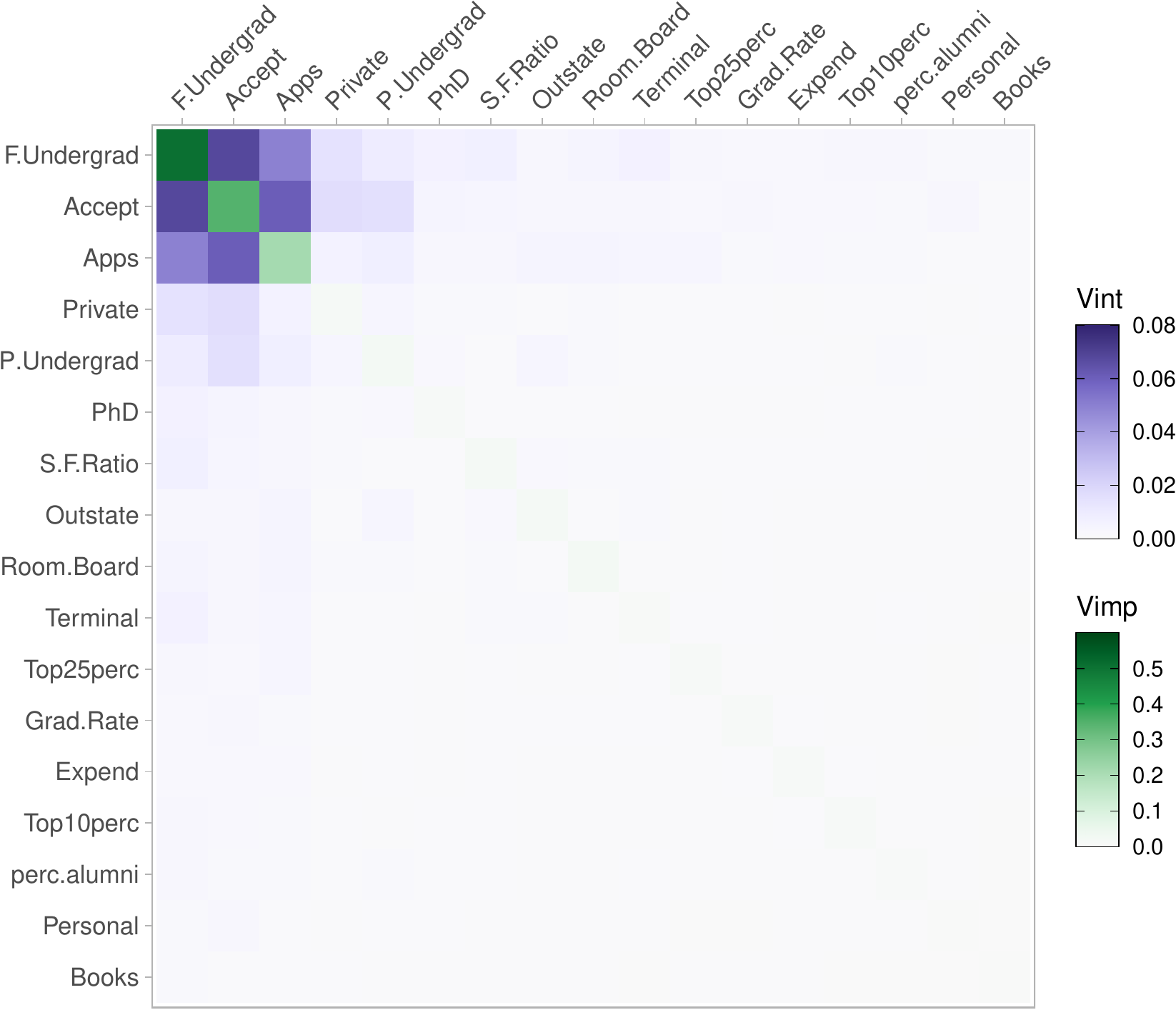} 
     \captionsetup{justification=centering}
     \caption{Random forest fit}\label{fig:figB}
   \end{subfigure}
\caption{A comparison of a kNN and random forest fit on the college application data. Both fits identify F.Undergrad as the most important variable as well as having similar mutual interactions between F.Undergrad, Accept and Apps. The kNN fit identifies many more moderate interactions between variables, especially concerning the variable Private} 
\label{fig:2plots}
\end{center}
\end{figure}

The heatmap display can be further used to compare different model fits. In Figure \ref{fig:2plots} 
we compare the random forest to a k-nearest neighbours (kNN) fit. 
In the left panel of Figure \ref{fig:2plots} we have a heatmap of a kNN fit (with $k = 7$ neighbours considered), while the right panel shows the random forest heatmap. To make a direct comparison of the heatmaps, we swap the embedded VImp measures that are available from a random forest fit and instead measure importance with an agnostic permutation approach that allows direct comparison of both the kNN and random forest models.  Furthermore, we set both heatmaps to use the same color scale for the VImp and VInt values.

We see in Figure \ref{fig:2plots} 
that both the random forest and kNN fit identify F.Undergrad as the most important variable for predicting the number of students enrolled. The top three variables are identical in both models, though the VImp values are much smaller in general across the kNN fit (e.g. the measured VImp for F.Undergrad for the kNN and random forest fits are 0.16 and 0.6 respectively). Both fits show mutual interactions between F.Undergrad, Accept and Apps. However, the kNN fit also suggests a moderate interaction between Private (i.e., whether the university was public or private) and F.Undergrad, which appears somewhat lower in the random forest fit. As Private has a relatively low VImp in both model fits, a simple VImp screening could miss its relevance to the fit. We note though, that this  kNN-random forest comparison  is for the sake of illustration only, as in this instance the kNN fits poorly by comparison with the random forest, having a test mean square error (MSE) over three times bigger. 

\subsection{Network visualization}
\label{sec:networkvis}

\begin{figure}[htbp]
\begin{center}
   \begin{subfigure}{0.4\linewidth} \centering
     \includegraphics[width=\linewidth]{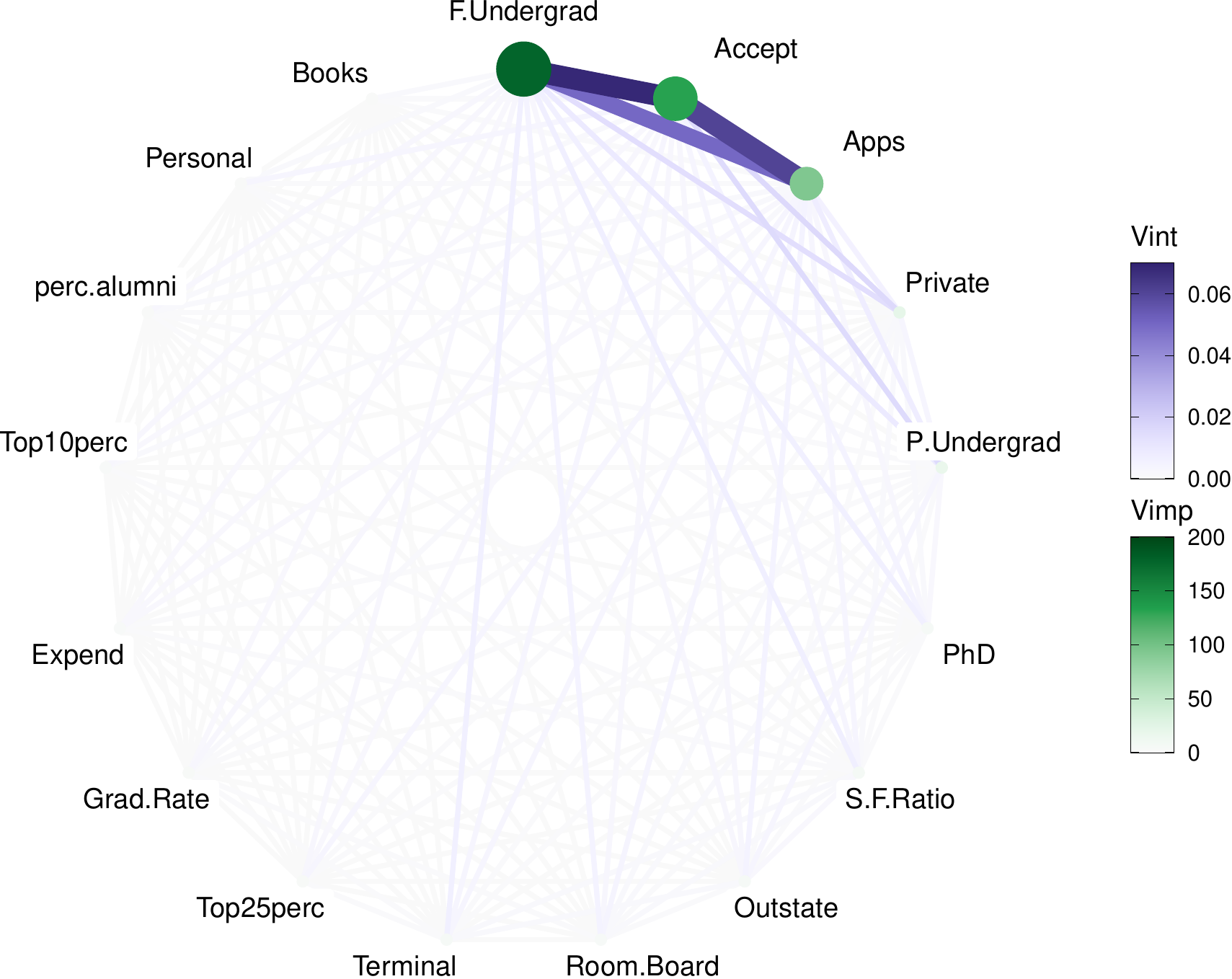}\vfill 
     \caption{}
     \label{fig:networkColl}
   \end{subfigure}
    \hspace*{.3cm}
   \begin{subfigure}{0.4\linewidth} \centering
     \includegraphics[width=\linewidth]{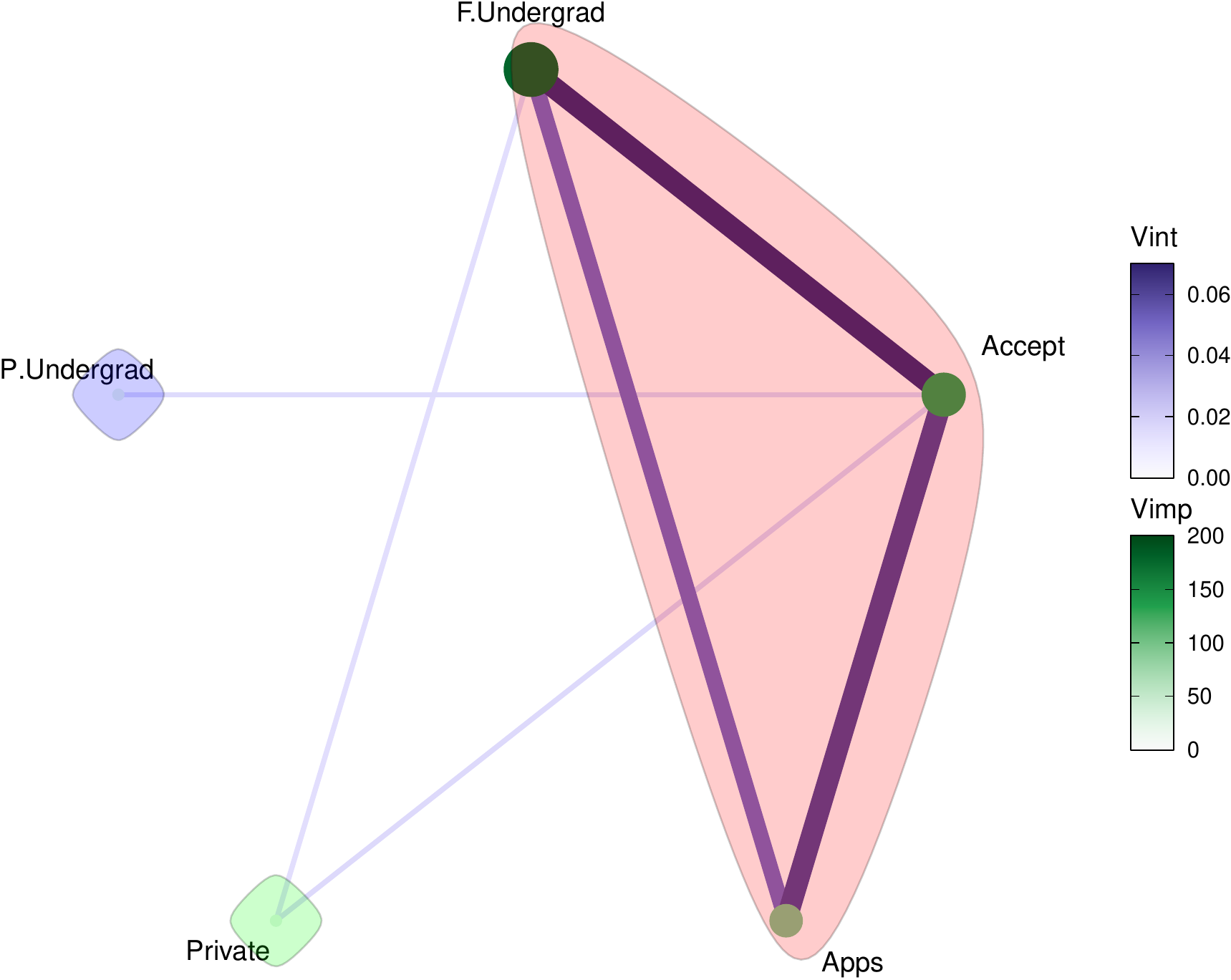} \vfill
     \caption{}
     \label{fig:networkCollT}
   \end{subfigure}
\caption{Network plot of a random forest fit on the college application data. Three mutually interacting and important variables can be seen, namely F.Undergrad, Accept and Apps. In (a) all of the variables are displayed. In (b) the network plot has been filtered to display pairs of variables with high VInt and clustered to highlight variables with mutually high VInt.} 
\label{fig:collNworks}
\end{center}
\end{figure}

As our second offering for displaying VIVI, we propose a network plot that shares similar benefits to the heatmap display but differs from it by giving a visual representation of the magnitude of the importance and interaction values not only via color but also by the size of the nodes and edges in a graph. 
In this plot, each variable is represented by a node and each pairwise interaction is represented by a connecting edge. See Figure \ref{fig:collNworks}(a) 
for an example. The color scales were chosen to match that used in the heatmap, with node size and color luminance increasing with variable importance. Similarly, edge width and color reflects the strength of the VInt. By default we choose a radial layout to display the variables (although this can be changed according to preference) and use the same seriation of variables as the heatmap, with the variables of high importance and high interaction strength placed in a clock-wise arrangement starting at the top. The benefit of such a display is that one can quickly decipher the magnitude of the importance and interactions of the variables  as well as seeing which variables both individually and jointly impact on the response.

In Figure \ref{fig:collNworks}(a) we again use the random forest fit of the college application data, using the same VImp and VInt measures as in Figure \ref{fig:dendser}.
In the network plot the strong mutual interactions between F.Undergrad, Accept and Apps and are represented by thick, intensely purple lines. F.Undergrad is identified as the most important single predictor and is represented by a large, intensely green node.
For settings with large number of predictors, it will be useful to filter the display to focus on high VIVI variables.
An additional step groups or clusters the variables according to VImp or VInt values.
For example, Figure \ref{fig:collNworks}(b) shows a network plot, filtered to display pairs of variables with high VInt and clustered to show groups
with mutually similar VInt.  
Here it is clear that the cluster colored pink  contains the variables with the largest VInt scores.  
In this example we use hierarchical clustering, but in our implementation, the graph clustering methods provided by the package \texttt{igraph} \citep{igraph} are  directly available.
 
 \section{Visualizing partial dependence and individual conditional expectation }
\label{sec:pdpVis}

We introduce new variants of partial dependence and individual conditional expectation plots in two different layouts.
With these plots, we can further investigate predictor effects singly and pairwise, especially for those predictors deemed important in our VIVI plots.
Additionally, our new plots combine displays of variable pairs, thus highlighting the presence of strong correlations where VInt measures may mislead. Conventionally, partial dependence plots are shown singly or in linear layouts, see Section 8.1 of \citet{molnar2019} for examples.  By comparison, our new displays are more compact, richer, and benefit from seriation.

\subsection{Individual conditional expectation curves}
\label{sec:pdp}
 \cite{ICEpaper} described individual conditional expectation (ICE) curves, which are closely related to partial dependence plots (PDPs). While a PDP shows the average partial relationship between the response and one or two features $S$, ICE plots  display a collection of curves, each showing the estimated relationship 
between the response and the feature $S$, at an observed value of  other features.
Recalling Equation (1), the ICE curves consist of $ g(\boldsymbol{x}_S, \boldsymbol{x}_{C_i} ) \; {\rm versus }  \; \boldsymbol{x}_S,  \; i= 1, 2, \ldots, n$, while the PDP curve is their average $f_S(\boldsymbol{x}_S)$.
If the ICE curves follow a similar pattern then the PDP is a useful overall summary, but if the pattern varies, then the feature effect is not homogeneous.

\subsection{Generalized partial dependence pairs plot with ICE curves}
\label{sec:GPDP}
\begin{figure}[htbp]
\centering
\includegraphics[scale = 0.6]{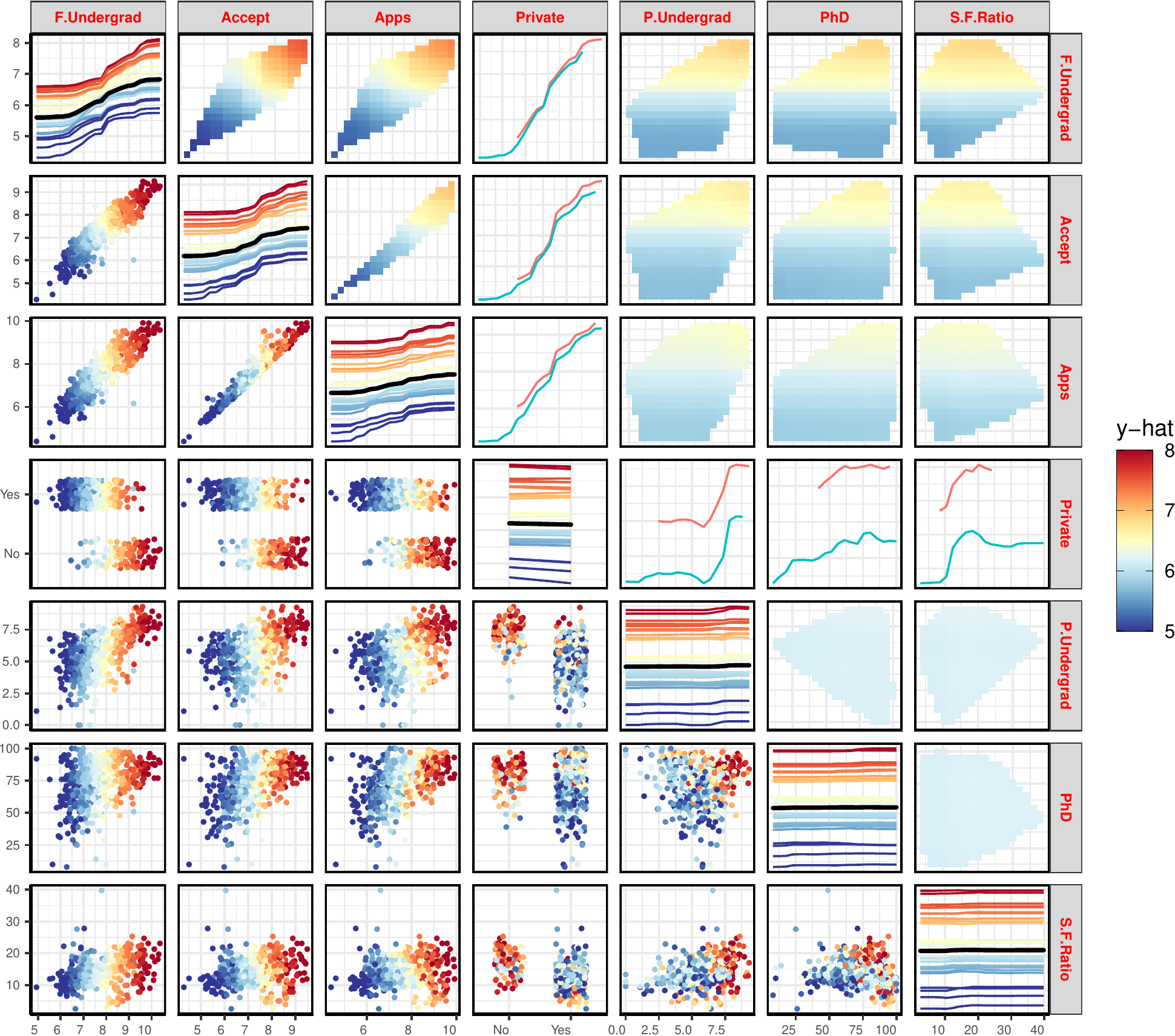}
\caption{GPDP of a random forest fit on the college data showing the seven most influential variables. From the changing one and two-way partial dependence, we can see that F.Undergrad, Accept and Apps have some impact on the response. However, as they are highly correlated and have similar increasing marginal effects, the potential interactions identified by the $H$-statistic are likely to be spurious.}
\label{fig:pdpColl}
\end{figure}

We  propose a generalized pairs partial dependence plot (GPDP) with one-way partial dependence and ICE curves with a superimposed partial dependence curve on the diagonal, the bivariate partial dependence on the upper diagonal and scatter plots of raw variable values on the lower diagonal, all of which are colored by the predicted values $\hat{y}$. Figure \ref{fig:pdpColl} 
provides an example. With the generalized pairs plot, an analyst can  quickly identify which variables singly or jointly impact on the fit. 
We use a diverging palette so deviations from the average response are emphasized. Here, high values of $\hat{y}$ are shown in dark red and low values are shown in dark blue. Mid-range values are shown in yellow. To avoid interpreting the PDPs where there are no data (and hence potentially spurious $H$-statistics), we mask out extrapolated areas by plotting the convex hull. 
For maximum resolution of the bivariate PDPs, the range of the collection of PDP surfaces dictates the limits of the color map. As predictions for individual observations and ice curves are likely to fall beyond these limits, colors are assigned using the closest value in the color map limits.

The ordering of the variables matches that of our heatmap and network plots. The GPDP differs from the previous plots by showing us the distribution of the explanatory variables (lower-diagonal), the exact nature of any linear/non-linear effects through the use of ICE curves (diagonal), and the average behaviour of the interactions through the use of two-way partial dependence (upper-diagonals). For the ICE curves we have limited the graphic to display a maximum of 30 randomly sampled curves by default, to allow individual ICE curves to be seen. As with the other visualizations, our GPDP can handle both categorical responses and predictors.

Figure \ref{fig:pdpColl} 
shows an example of a GPDP of the college applications data. In the interest of space, we pre-filter this plot to show  the seven most influential variables. The bivariate PDPs show the response surface over the convex hull of each variable pair. The lower diagonal plots indicate that F.Undergrad:Accept, F.Undergrad:Apps  and especially Accept:Apps are highly correlated with similar increasing marginal effects on the diagonal,
 suggesting that the the high $H$-values between these variables are likely to be spurious. This is verified by the bivariate linear PDPs for these variables.
As Private is a factor with two levels (i.e., yes or no), the partial dependence for each factor level is shown in the upper-diagonal (with yes in red and no in blue). The remaining variables would appear to have little effect either singly or jointly on the response. This can be seen from the flat one-way PDP and ICE curves on the diagonal and the flat  two-way PDPs.

\subsection{Partial dependence zenplot}

Our final display uses the methods of \citet{ZEN} to show selected panels of the all-pairs PDP in a space-saving layout, which we call a zen-partial dependence plot (ZPDP). Zenplots (zigzag expanded navigation plots) were designed for showing pairwise plots of high-dimensional data in a zigzag layout.
The motivation for zenplots  is that they focus on interesting 2D displays, and they permit examination of high-dimensional data.
Indeed, \citet{HofertOldford2018} present an example where they successfully explore pairwise dependence of 465 variables via 164 zenplots. 
Here we propose to adapt zenplots for bivariate partial dependence plots. 

To describe the construction,
consider a network plot showing VImp/VInt such as that in Figure \ref{fig:collNworks}(a). 
Then delete edges with VInt below a threshold, leaving a graph such as that in Figure \ref{fig:collNworks}(b). We wish to build partial dependence plots showing pairs of variables with high VInt, that is, visiting each of the edges in our thresholded graph. For a connected graph, the greedy Eulerian path algorithm of \citet{PairViz1} visits each edge at least once, starting from the highest weighted edge and moving through edges giving preference to the highest-weight available edge. If the graph is not even, some edges may be visited more than once, or additional edges are visited. If the graph is not connected, we form sequences for the connected sub-graphs, which are optionally joined into a single sequence. 

Zenplots use the zigzag display algorithm of \cite{ZEN} and allow for the display of high-dimensional data by alternating plot axes in a zigzag-like pattern where adjacent axes share the same variable. We adapt this concept replacing bivariate data plots with bivariate partial dependence plots. As interpretation issues may arise when the distribution of some of the variables is highly skewed, we display a rug plot on each axis to show the distribution of the data. For ease of viewing, the rug plots are a single color and use alpha blending to highlight the distribution.
As with our GPDP, there is an option to mask areas where the partial dependence has been extrapolated. 
The resulting plot displays the most important interacting variables in as small a space as is possible, vastly reducing the number of plots that would be required for interpretation compared with a default matrix scatter plot of PDPs. 

\begin{figure}[htbp]
\centering
\includegraphics[scale = 0.45]{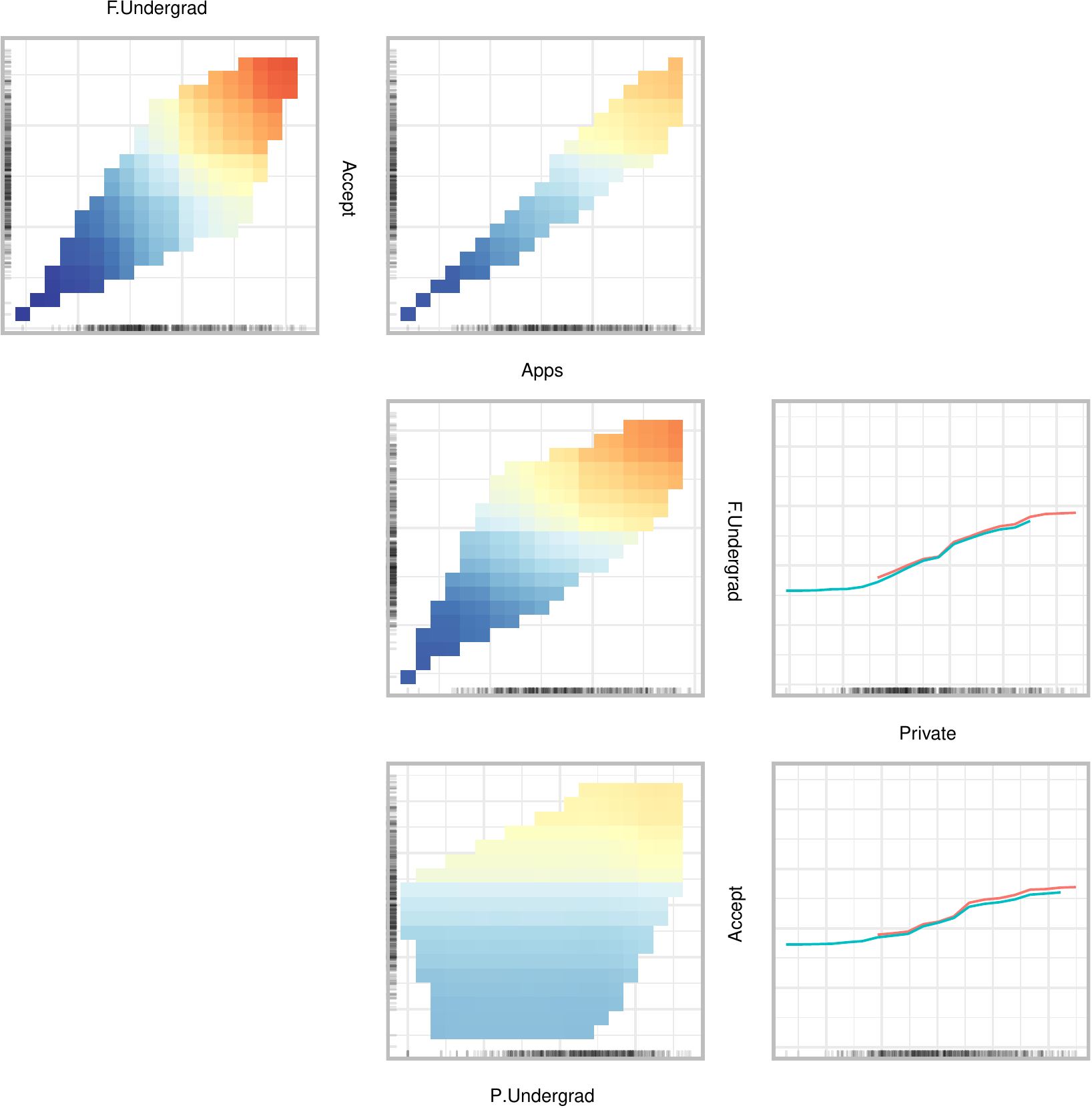}
\caption{ZPDP of a random forest fit on the college data. We can see that the predicted value for the number of students enrolled increases with each of the variables.}
\label{fig:collZen}
\end{figure}

In Figure \ref{fig:collZen}, 
we show a ZPDP for the random forest fit to the college applications data. 
The ZPDP shows the bivariate PDPs corresponding to each of the edges of Figure \ref{fig:networkCollT}.
The sequence of plots is obtained from an Eulerian visiting the edges starting with the highest-weight edge, here that is between
F.Undergrad and  Accept, and following available edges in order of preference by weight thereafter.
The resulting Eulerian is
F.Undergrad, Accept,   Apps,  F.Undergrad, Private,     Accept,    P.Undergrad.
The plots shown correspond to a subset of those in Figure \ref{fig:pdpColl}, limited to the
more interesting high-interaction pairs. This more compact display helps focus the reader's attention where it is needed, especially as the plots
are approximately ordered by decreasing $H$-index. 
The variables Private and P.Undergrad show little evidence of marginal importance, and 
notwithstanding the relatively large $H$-values, there is not much evidence of interaction with other predictors.

 \section{Case study: cervical cancer risk classification}
\label{sec:example}

Cervical cancer remains one of the most prevalent forms of cancer in women globally, ranking fourth in the global cancer incidence in women \citep{cancerRates}. 
The link between cervical cancer and sexually transmitted diseases (STDs) has been well established.
The long-term use of hormonal oral contraceptives is associated with increased risk \citep{OCrisk}.
Furthermore, having multiple children has been shown to increase risk \citep{NumberKidsRisk}, particularly in women previously infected with HPV.

Here we examine and create visualizations for data concerning cervical cancer risk factors \citep{cervicalData}. 
Based on the previous studies, we would expect our visualizations to align with prior identification of important variables, with the addition of gaining new information about how the variables interact. 
The data is comprised of historical medical records (such as a patient's STD history, oral contraceptive or intrauterine device [IUD] use) and personal information (such as age and sexual activity). Due to the personal nature of the questions asked for the collection of the data, several patients decided not to answer some of the questions, particularly those concerning  STDs. The data has been previously studied \citep[for example see][]{cervicalStudy3}. The full dataset contains 36 variables with 858 observations and uses Biopsy (Healthy or Cancer)  as the response. 

For this case study, we use a subset of the variables (see supplementary materials for a listing). Preliminary exploration of the data shows that many variables are highly skewed and contain zero values; in this case we use a $\log(x +1)$ transformation.
The data is split 70-30 into training and test sets. We fit a classification gradient boosting machine (GBM) model \citep{Friedpdp} to the training data, with Biopsy as the response.   The accuracy on the test set was measured to be 0.93, and the area under the curve (AUC) was 0.73. All plots were made using the training data, with all PDPs and the $H$-statistic measured on the logit scale.

\begin{figure}[htbp]
 \centering
 \includegraphics[scale = 0.5]{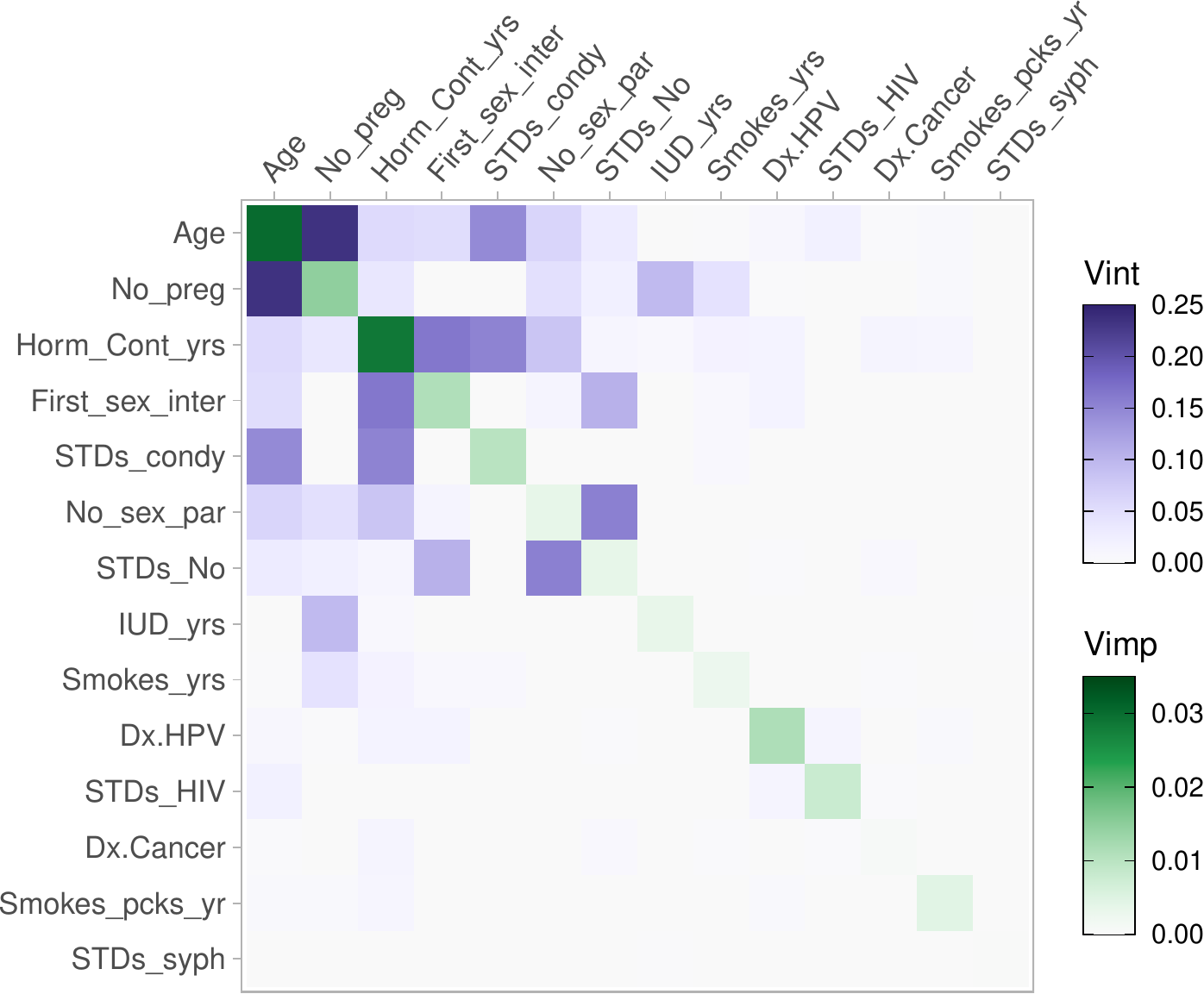}\vfill
  \caption{Heatmap of a GBM fit on the cervical cancer  data. 
  The first seven variables have the highest VIVI scores.
  Age and No\_preg have the strongest interaction.
  }
    \label{fig:cervicalHeatFull}
\end{figure}

Figure \ref{fig:cervicalHeatFull} 
displays a heatmap of the GBM fit on the cervical cancer risk data, using a permutation VImp method.
Reading from the top-left, the first seven variables have the highest VIVI scores. Overall, Age has the highest importance  followed closely by Horm\_Cont\_yrs (the number of years a patient has taken hormonal contraceptives). This is in agreement with the  studies mentioned above.
 Age also shares the strongest interaction with No\_preg (number of pregnancies), which has a medium Vimp but is highly important in terms of its interaction. We can see multiple interactions throughout the top seven variables. Of note is the interaction between STDs\_No (number of STDs a patient has previously had) and No\_sex\_par (number of sexual partners). Both of these variables share a strong interaction but have low VImps and they may have been mistakenly eliminated from a model were
VImp scores to be used as the sole variable selection metric.

We further explore the impact of the top five variables from Figure \ref{fig:cervicalHeatFull}
\begin{figure}[ht]
\centering
\includegraphics[scale = 0.8]{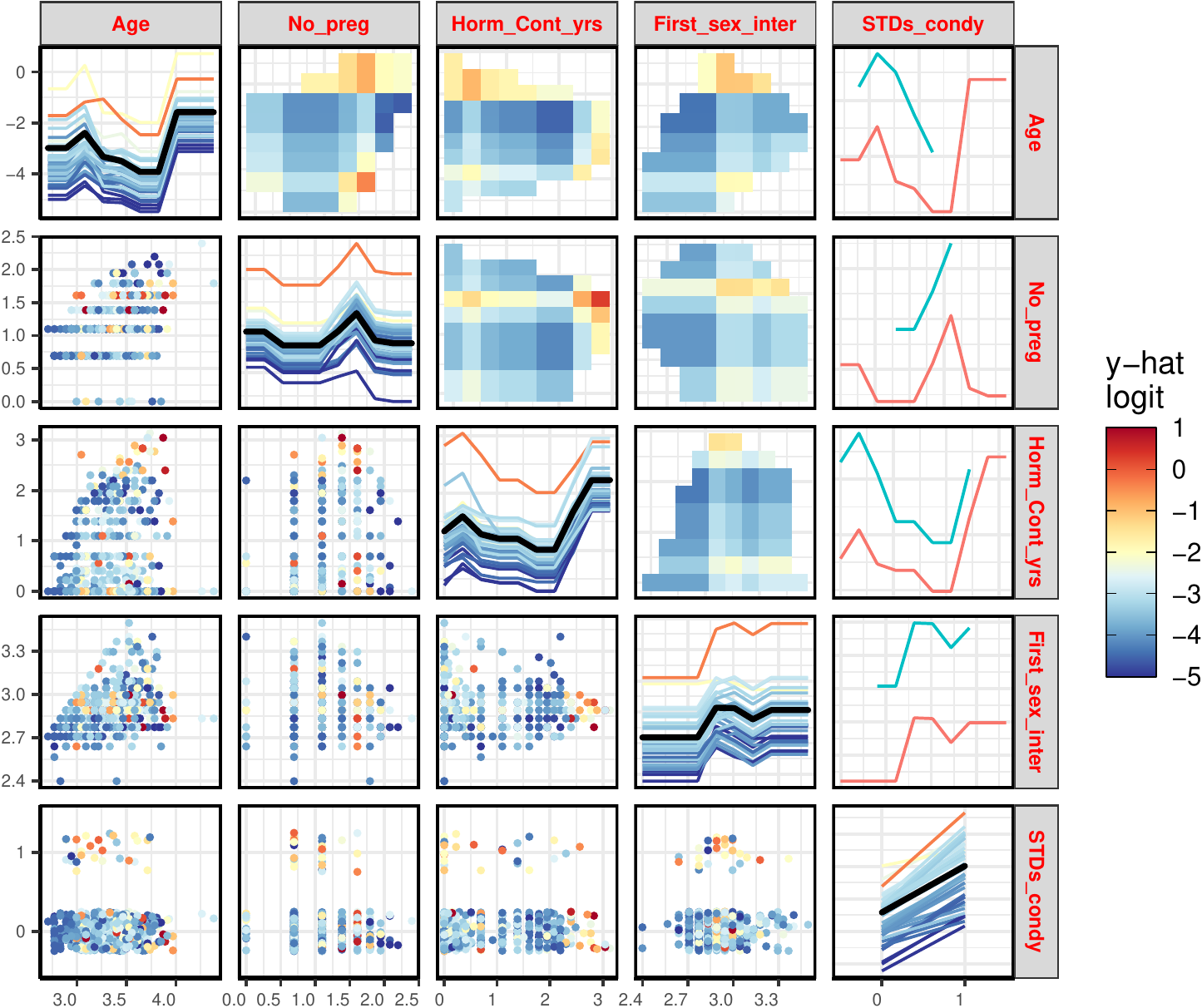}\vfill
\caption{GPDP of a GBM fit on  the cervical cancer data.
The presence of the STD condylomatosis (STDS\_condy in turquoise) increases the risk of cervical cancer. Risk increases substantially at higher ages and with prolonged use of hormonal contraceptives.
}
\label{fig:cervicalPDP}
\end{figure}
on cancer classification in the  GPDP plot of
Figure \ref{fig:cervicalPDP}. To compare response groups, the ICE plots on the diagonal show 25 instances sampled from each of the Cancer and Health groups. 
The ICE curves are colored according to the predicted log-odds of cancer for that instance. As there is only one red curve, the predicted model accords most observations low
cancer probabilities, even for those known to have cancer.
The solid black lines on the diagonal of Figure \ref{fig:cervicalPDP} show single variable PDPs.
The PDP curve for Age, the single most important predictor, has a mostly decreasing log-odds trend up to an age of 43 ($\approx$  3.75 on the $\log$ scale), with a steep incline thereafter.
But we can see from the Age scatterplots there are few cases with ages beyond 43, so the pattern in this area is not supported by much data.
The pattern for the Horm\_Cont\_yrs PDP is similar to that for Age, where log-odds of cervical cancer increases rapidly beyond eight years. In this case though, there are
quite a few observations in this region supporting this finding.

According to Figure \ref{fig:cervicalHeatFull}, the predictors No\_preg and Age have the strongest interaction. The bivariate PDP plot for No\_preg:Age
indicates the form of this interaction. A high number of pregnancies is associated with low cancer probability for middle age groups, but is associated with a higher cancer probability for older and, interestingly, younger patients. 
Note that in the plots with one numeric and one categorical variable, such as the plot for STD\_condy (STDs: condylomatosis) and Age, the numeric variable is always drawn on the x-axis, 
notwithstanding the label is on the y-axis. This is to allow the plot to be more easily read. In this plot, the bivariate PDP is the same as two PDPs for each level of STDs\_condy (where the green curve is for STDs\_condy $= 1$). Although this pair has a relatively high VInt score (as seen in Figure \ref{fig:cervicalHeatFull}), there does not appear to be an interaction present in the bivariate PDP, as the difference between the two curves does not vary with age.

\begin{figure}[htbp]
\centering
\includegraphics[scale = 0.55]{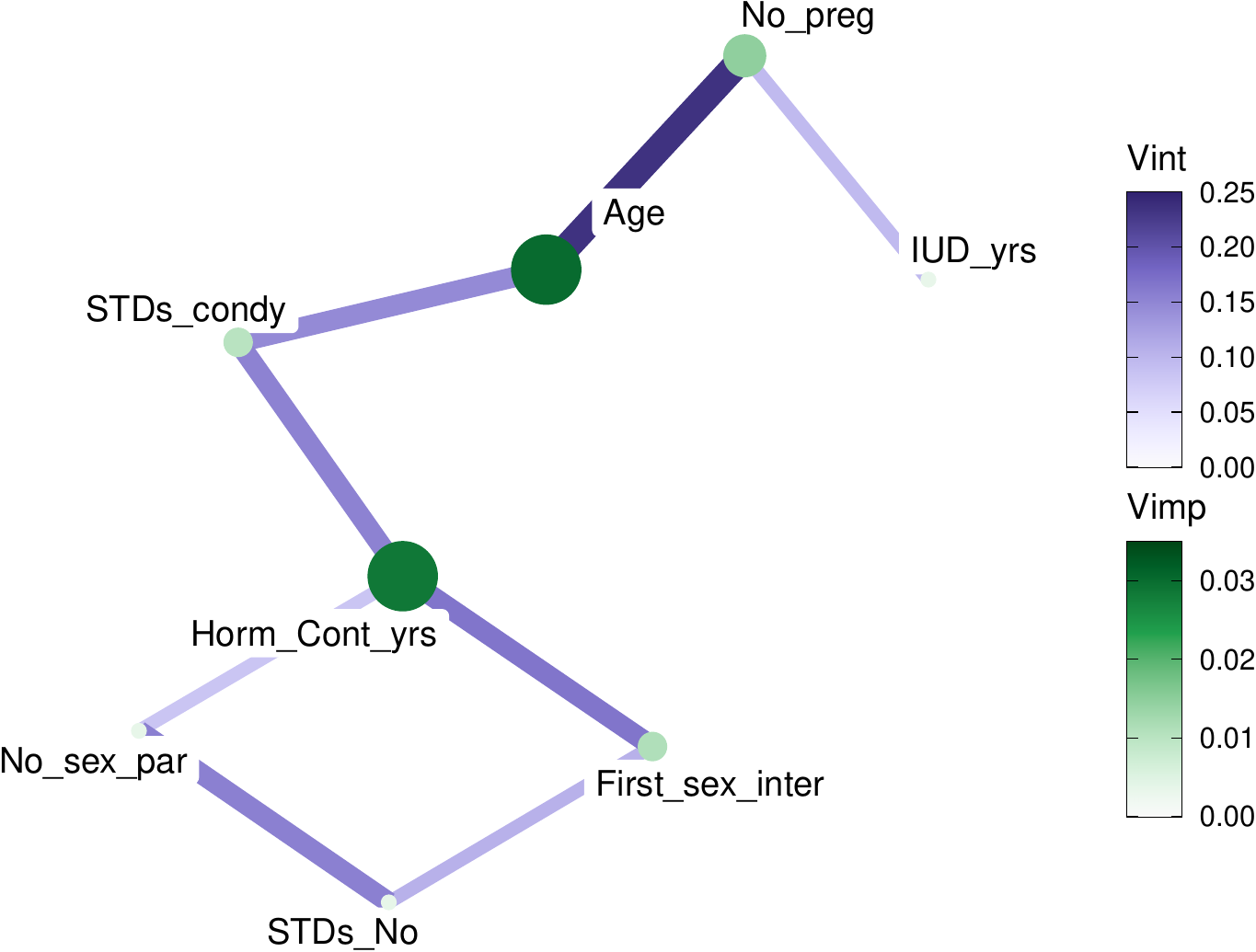}
\caption{Network graph of a GBM fit on the cervical cancer  data, filtered to show pairs of variables with $H$-index greater than 0.08. }
\label{fig:cervicalNet}
\end{figure}

To focus just on predictors with high pairwise interaction scores, we turn to a network plot. 
Figure \ref{fig:cervicalNet} 
displays a network plot of the GBM fit to the cervical cancer risk data,
filtered to show pairs of variables with a $H$-index greater than 0.08 (with the cutoff chosen after inspection of the histogram of $H$ values). The selected variables include the five variables appearing in Figure \ref{fig:cervicalPDP}, and
three additional variables, namely No\_sex\_par, STDs\_No, and IUD\_yrs (number of years with an intrauterine device), with eight relevant interactions
between them. This display has some benefits over the heatmap display of Figure 
\ref{fig:cervicalHeatFull}. Firstly,  it focuses directly on pairs of variables with high interaction, particularly with the choice of network layout.
Secondly, in the heatmap plot, even with seriation, some high-interaction pairs of variables may not be positioned nearby which detracts from readability. For example in Figure \ref{fig:cervicalHeatFull},  associating the relevant variables
with the strong interaction for (First\_sex\_inter, 
STDS\_No) requires considerable effort from the reader. However, this strong interaction is immediately obvious in Figure \ref{fig:cervicalNet}.

\begin{figure}[ht]
\centering
\includegraphics[scale = 0.6]{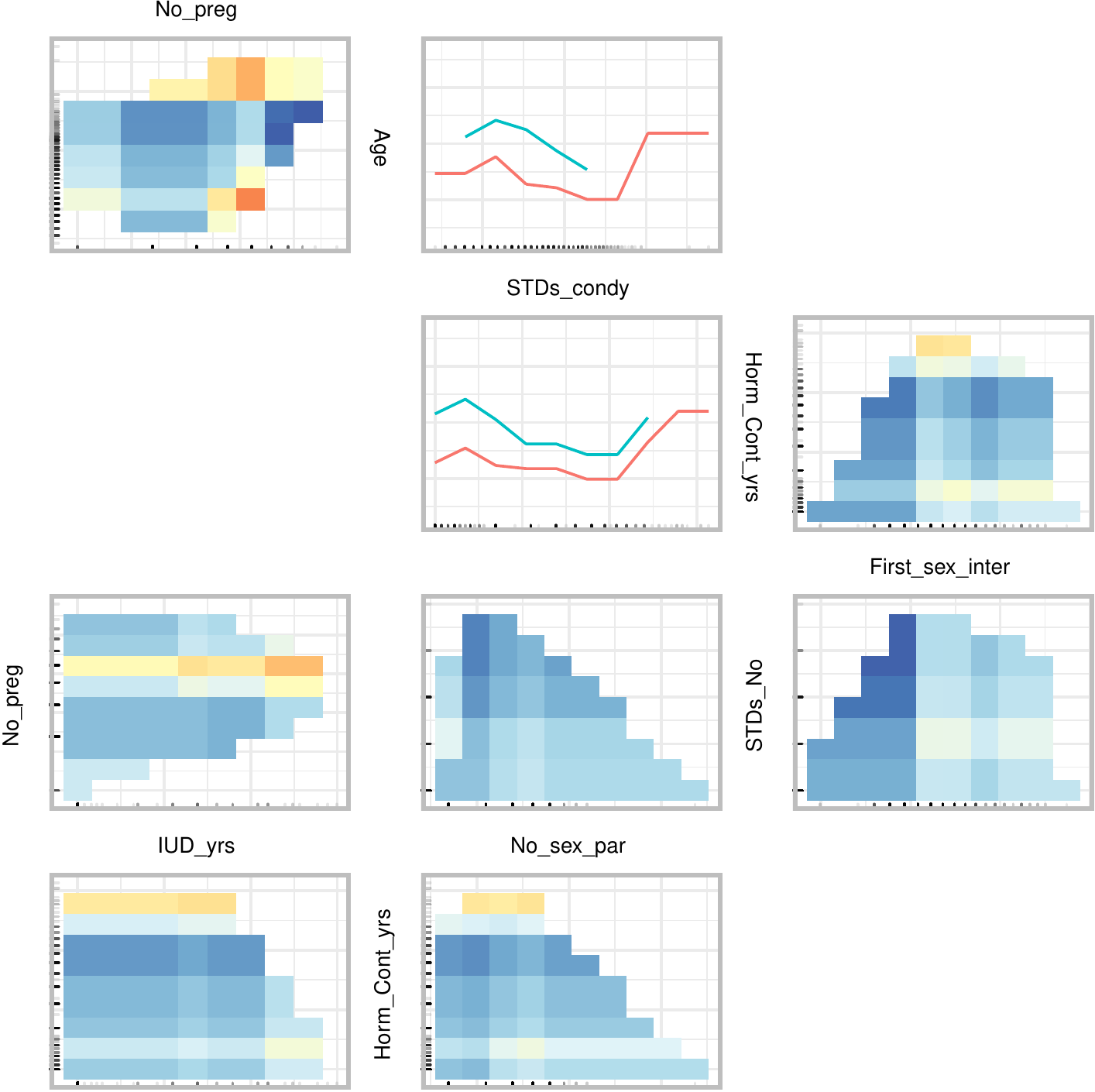}
\caption{ZPDP of a GBM fit on the cervical cancer risk data. High cancer probability occurs with  high number of pregnancies and high age.
The color scale is the same as that of  Figure \ref{fig:cervicalPDP}.}
\label{fig:ZPDP}
\end{figure}

To explore these interacting variables further, we use a ZPDP
in Figure \ref{fig:ZPDP} 
to show the bivariate PDPS for the eight interactions. The Eulerian path
starts with the pair of variables with the highest $H$-index (here No\_preg:Age), and from there to Age:STDs\_cond\_y, ending up at No\_preg:IUD\_yrs.
An additional plot is added corresponding to an edge between Horm\_Cont\_yrs and IUD\_yrs to complete the Eulerian. (In this example, it would be
possible to construct a ZPDP based on an Eulerian visiting each edge of the graph in Figure \ref{fig:cervicalNet} exactly once, but this Eulerian
ignores edge weights.)
The STDs\_No:No\_sex\_par plot (third row, second column) is a flat surface with no evidence of interaction, despite these variables having a moderate $H$-index.
Interestingly, in the No\_preg:IUD\_yrs plot (third row, first column), the probability of developing cervical cancer is increasing with  IUD\_yrs, with a steeper gradient for moderately high No\_preg. 
Further investigation is needed to determine the nature of this effect.

To summarize, we have used our visualizations to identify and examine some clear risk factors associated with developing cervical cancer. Our novel approach allowed us to examine specific pairs of variables that interact and through our use of graphs and PDPs, we can examine how each variable affects the model's predictions.  Specifically, the age of a patient and the number of years of hormonal contraceptive use seem to be important risk factors,
agreeing with previous studies.
From Figure \ref{fig:cervicalPDP}, the women who took hormonal contraceptives for eight or more years appear to have a higher risk cervical cancer,
which is in agreement with the findings of \cite{OCrisk}.
Surprisingly, as seen in Figure \ref{fig:cervicalHeatFull}, Dx.HPV (i.e., whether the patient has had a previous diagnosis of HPV) was ranked to have middling importance, despite the known link between HPV and cervical cancer. 
Neither did we see evidence of an interaction between No\_preg and Dx.HPV, which contrasts with \citet{NumberKidsRisk}.
These differences may be due to the low frequency of positive cases in the data. 

\section{Discussion}
\label{sec:conclusion}

We have presented  innovative and informative methods to visualize the importance and interactions of variables simultaneously from a model. 
The seriated heatmap of Section \ref{sec:heatmapvis} and the network plot of Section \ref{sec:networkvis} 
are effective in determining which variables have the most impact on the response in a model fit.
We view VIVI measures displayed in heatmap and network plots as a starting point for further detailed exploration of the nature of variable
effects and interactions in the GPDP and ZPDP.
The ZPDP construction is a novel application of the recently proposed zenplots of  \cite{ZEN}, which should prove particularly useful to
focus exploration on high-VIVI subsets of variables. 


Our methods are intuitive, flexible and easily customisable.
Built-in or model-agnostic variable importance measures may be used in our heatmap and network displays.
In our work to date, we use the model-agnostic $H$-statistic. 
Model-agnostic measures are particularly useful when comparing two or more fits. The heatmap and network displays will also be useful for comparing  different VIVI measures
for the same fit.

As calculation of the VIVI matrix and  our  visualizations  are available for any subset of the data, stratified versions or facetted displays will give insight into higher-order predictor interactions. 
A drawback to the $H$-statistic is  calculation speed, which is highly model dependent, though sampling and parallel calculation offer useful speed-ups.
For example, the 14 $\times$ 14 $H$-matrix for the GBM fit in Figure \ref{fig:cervicalHeatFull}  computed on  30 randomly selected observations
took approximately 16 seconds on a MacBook Pro 2.3 GHz Dual-Core Intel Core i5 with 8GB of RAM. 
Calculation for the 17-predictor random forest fit in  Figure \ref{fig:dendser} is much slower, taking approximately 79 seconds, even though
here we used just 20 randomly selected observations.
A second drawback we have identified is that high $H$ values can occur in settings where there is no feature interaction, especially in the
presence of high variable correlation.
The presence and nature of interactions can be further verified in the bivariate partial dependence plot, thus avoiding misleading conclusions.

 A bivariate importance measure, perhaps obtained by permuting pairs of variables, could be used in place of the $H$-statistic in the heatmap and network visualizations. 
It would also be interesting to explore the interaction measures of \citet{VIN} and \citet{vip} in our visualizations, and whether these measures avoid the issues identified with the use of $H$.

A number of variants of the GPDP and ZPDP could be investigated in future work.
One possibility for the bivariate PDP,  is to subtract  the two marginals plotting $f_{jk} - f_{j} - f_k$, which corresponds
directly to the $H$-statistic.
Alternatively, accumulated local effects (ALE) functions \citep{ALE} could be used in place of PDPs in our matrix layouts.
ALE functions were constructed with the goal of counteracting the bias issues of partial dependence functions. 
Another option might be to replace the partial dependence $f$ in Equation  \ref{eqn:HStatInPapernn} with the corresponding ALE function,
giving a new interaction measure.

\section*{APPENDIX}


We explore some limitations of the $H$-statistic using a simulated dataset.
We demonstrate the benefits of the un-normalized version of $H$, and  show how correlated variables can result in spuriously high interaction measures.

Using the Friedman benchmark equation \citep{FriedmanEqn},
\begin{align}
\label{friedData}
y = 10 \sin(\pi x_1 x_2) + 20 (x_3 - 0.5)^2 + 10 x_4 + 5 x_5 + \epsilon \\
{\rm where} \;\; x_j \sim  U(0, 1), j=1,2,\ldots, 10;\;\;   \epsilon \sim N(0, 1) \notag
\end{align}
we simulate 1,000 observations and fit a random forest.
There are five important variables with an interaction between $x_1$ and $x_2$,
and five additional predictors $x_6, x_7, \ldots x_{10}$ unrelated to the response. 

\begin{figure}[htbp]
\begin{center}
   \begin{subfigure}{0.3\linewidth} \centering
     \includegraphics[scale=0.25,trim={0 0 40 0},clip]{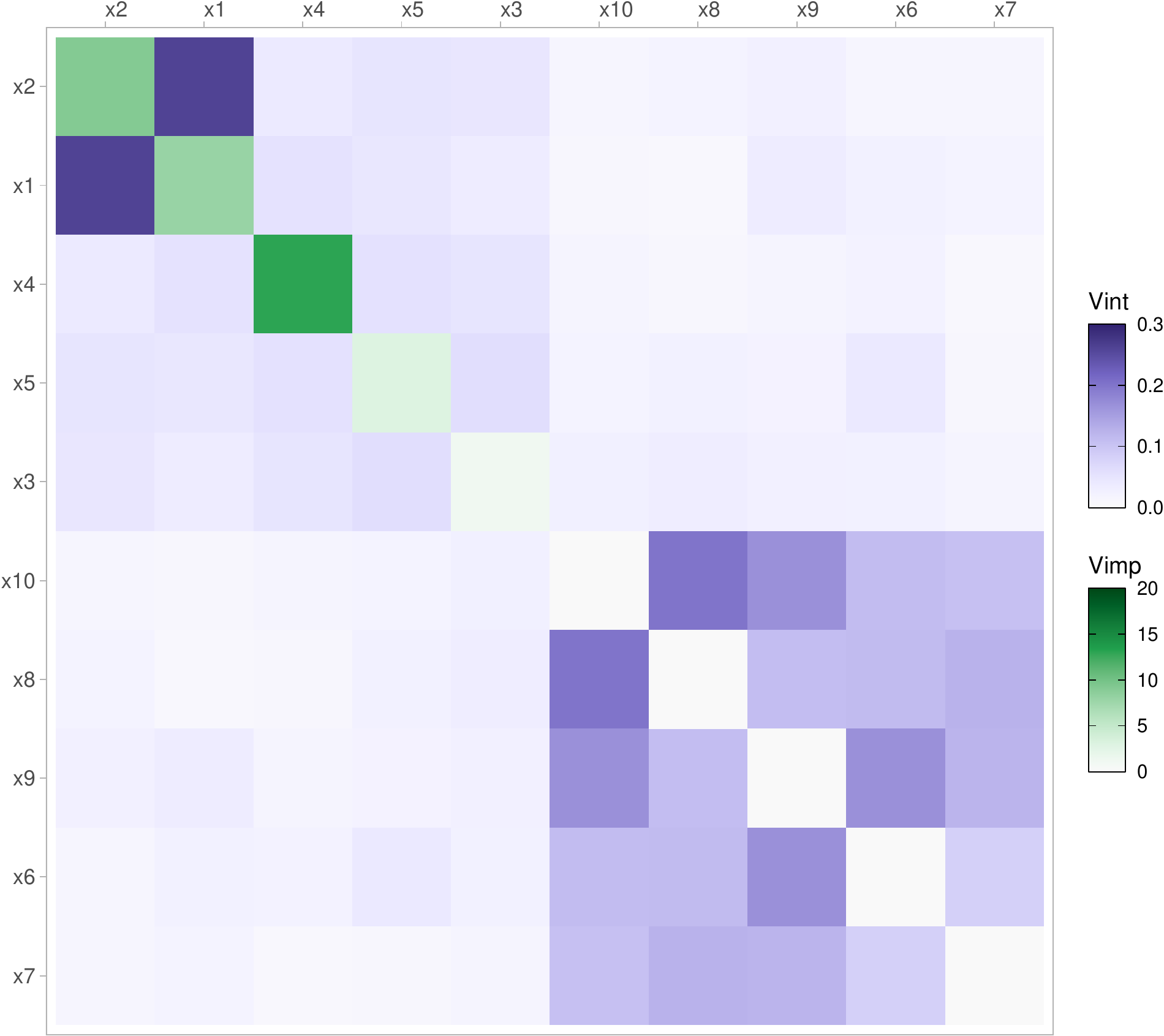} 
     \captionsetup{justification=centering}
     \caption{Normalized $H$}
   \end{subfigure}
   \begin{subfigure}{0.3\linewidth} \centering
     \includegraphics[scale=0.25,trim={0 0 40 0},clip]{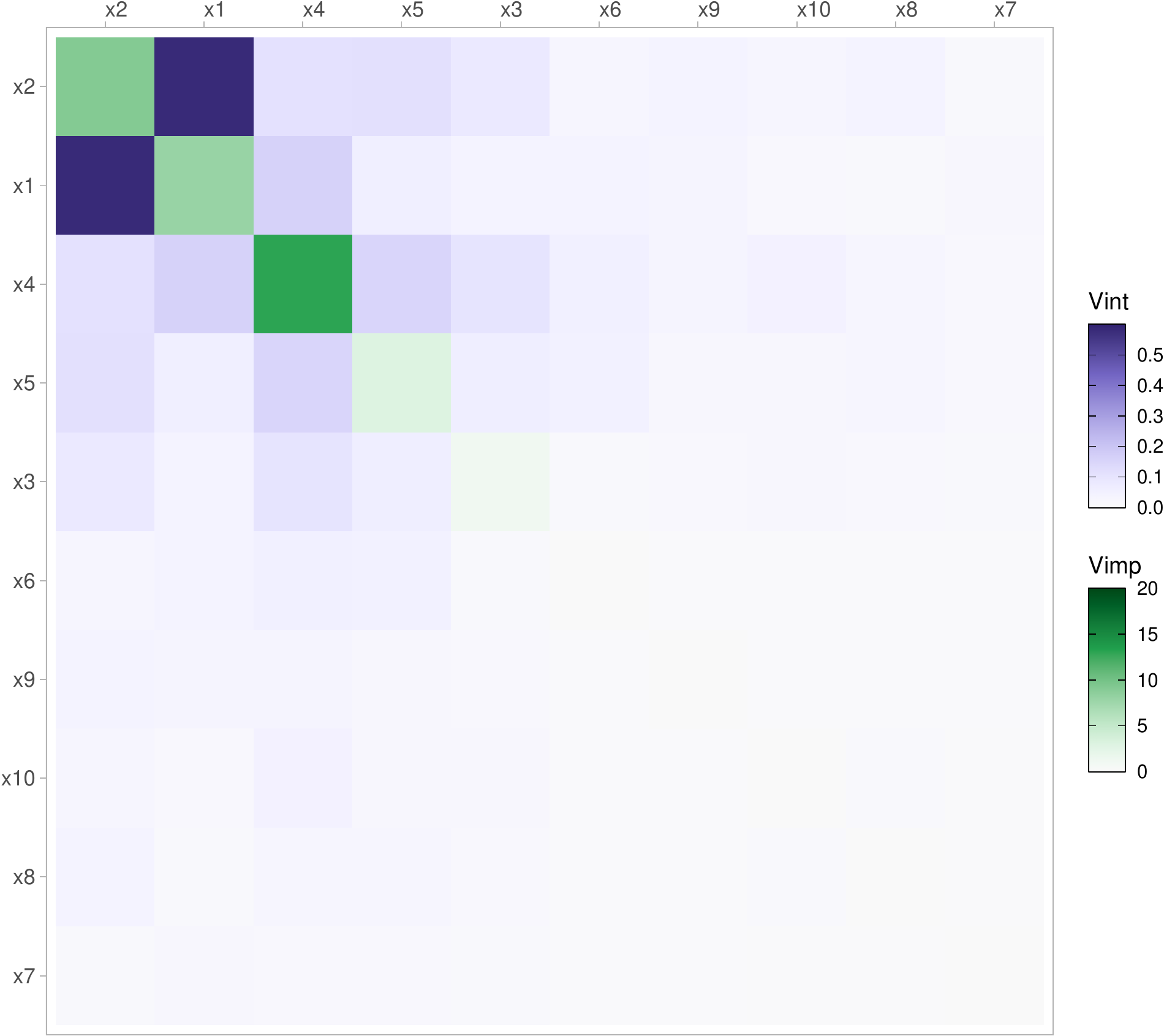} 
     \captionsetup{justification=centering}
     \caption{Un-normalized $H$}
   \end{subfigure}
   \begin{subfigure}{0.3\linewidth} \centering
     \includegraphics[scale=0.25,trim={0 0 40 0},clip]{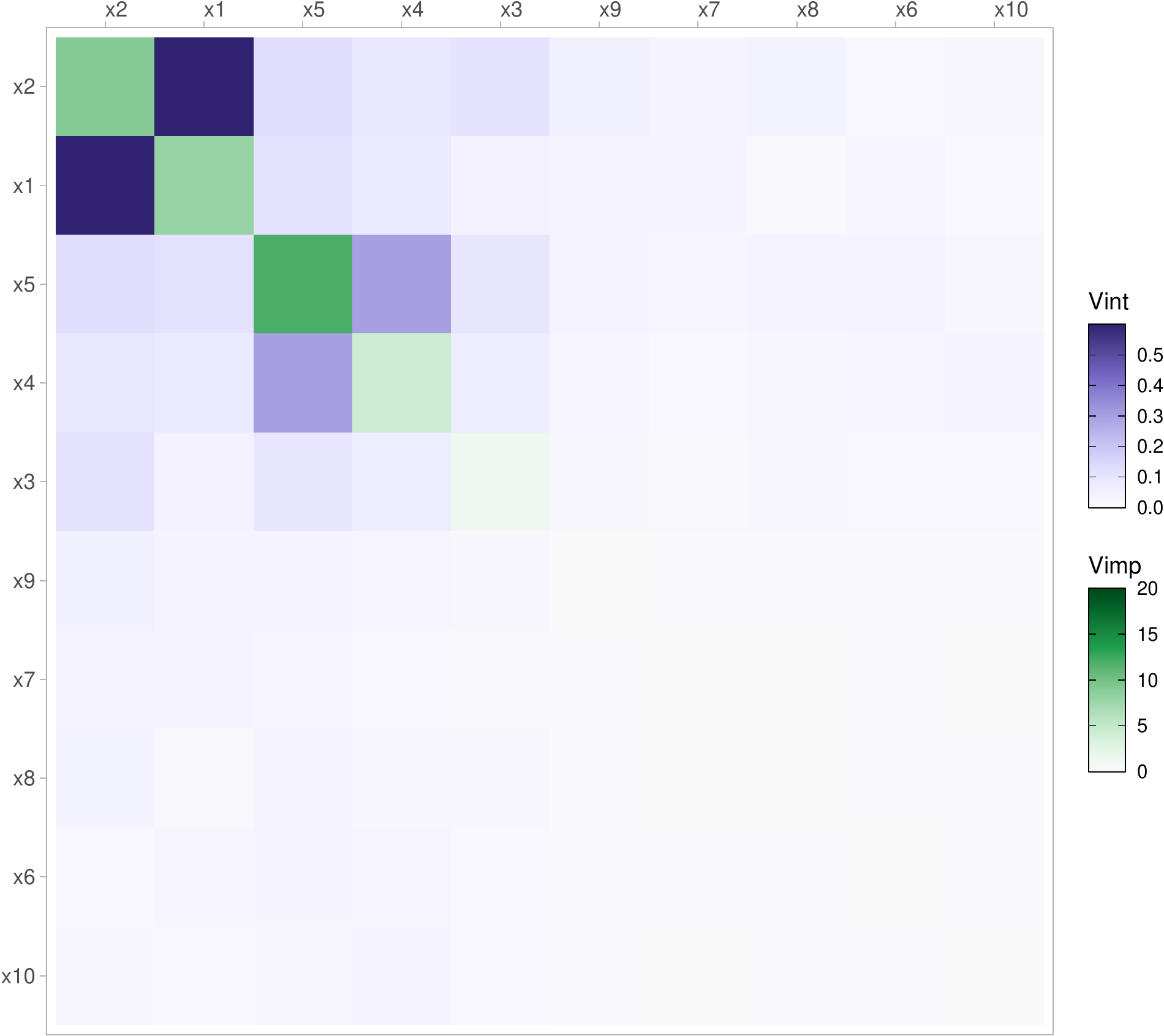} 
     \captionsetup{justification=centering}
     \caption{Correlated variables.}
   \end{subfigure}
\caption{Comparison of the normalized and un-normalized $H$-statistic and the effect of including correlated variables for a random forest model. In (a) multiple spurious interactions are detected when using the normalized  $H$-statistic. In (b) the spurious interactions have mostly disappeared when using the un-normalized version. In (c) (un-normalized $H$) a moderate spurious interaction between the correlated variables $x_4$ and $x_5$ is detected .}
\label{fig:spurious}
\end{center}
\end{figure}

In Figure \ref{fig:spurious}(a) and (b)
we compare the normalized and un-normalized versions of the $H$-statistic for the simulated data. 
Colour legends are not useful here and are omitted.
In all cases, the $x_1$:$x_2$ interaction is correctly identified.
However, in (a) there are numerous spurious strong interactions among the noise variables. 
In (b)  using  un-normalized $H$ these spurious strong interactions disappear.
The culprit here is the denominator in Equation \ref{eqn:HStatInPaper}, which for variables $x_6, x_7, \ldots x_{10}$  will be close to zero, thus artificially inflating $H$.
This is the rationale behind our use of the un-normalized $H$-statistic in our examples throughout.

 A more subtle  cause of spuriously inflated $H$ is due to bias in the partial dependence curve. 
 This is a particular problem in the presence of correlated predictor variables \citep[for example, see][]{ALE}.
 To demonstrate this, we replace $x_5$   with $0.3x_5 + 0.7x_4$ in Equation \ref{friedData} thus introducing a strong correlation ($\approx 0.92$) between $x_4$ and $x_5$.
The resulting VIVI heatmap of the random forest fit in Figure \ref{fig:spurious}(c)
shows a moderate $x_4$:$x_5$ interaction which is spurious.
Even  in the absence of correlation, bias can occur  if the fit exhibits regression to the mean.
For example, this occurs with tree-based fits such as a random forest, where  predictions cannot lie outside the range of training set responses.
This bias is evident in Figure \ref{fig:spurious}(b) and (c) as the light purple squares in the top-left section of the heatmaps.

\bigskip
\begin{center}
{\large\bf SUPPLEMENTARY MATERIAL}
\end{center}

\begin{description}

\item[Datasets] description in datasets.pdf

\item[College] Code for examples of  Section \ref{sec:vimpvint} and \ref{sec:pdpVis}
in  college.Rmd

\item[Cervical] Code  for examples of Section  \ref{sec:example} 
 in  cervical.Rmd

\item[hstat] Code for example of Appendix   in hstat.Rmd

\end{description}

\bibliographystyle{asa}

\bibliography{vivid.bib}

\end{document}